\documentclass[11pt,prd]{revtex4-1}
\usepackage{graphicx, amsmath, amssymb, color}
\usepackage{bm}
\usepackage{hyperref}
\usepackage{graphicx}
\usepackage{subfigure}

\def\aap{Astronomy \& Astrophysics}

\begin{document}
\title{Forecasting Cosmological Bias due to Local Gravitational Redshift}
\author{Haoting Xu}
\affiliation{School of Physics and Astronomy, Sun Yat-sen University, 2 Daxue Road, Tangjia, Zhuhai, China}
\author{Zhiqi Huang}
\email{huangzhq25@mail.sysu.edu.cn}
\affiliation{School of Physics and Astronomy, Sun Yat-sen University, 2 Daxue Road, Tangjia, Zhuhai, China}
\author{Na Zhang}
\affiliation{School of Physics and Astronomy, Sun Yat-sen University, 2 Daxue Road, Tangjia, Zhuhai, China}
\author{Yundong Jiang}
\affiliation{School of Physics and Astronomy, Sun Yat-sen University, 2 Daxue Road, Tangjia, Zhuhai, China}
\date{\today}
\begin{abstract} 
  When photons from distant galaxies and stars pass through our neighboring environment, the wavelengths of the photons would be shifted by our local gravitational potential. This local gravitational redshift effect can potentially have an impact on the measurement of cosmological distance-redshift relation.  Using available supernovae data, Ref.~\cite{Local} found seemingly large biases of cosmological parameters for some extended models (non-flat $\Lambda$CDM, $w$CDM, etc.). Ref.~\cite{Local_H} pointed out that, however, the biases can be reduced to a negligible level if cosmic microwave background (CMB) data are added to break the strong degeneracy between parameters in the extended models. In this article we forecast the cosmological bias due to local gravitational redshifts for a future WFIRST-like supernovae survey. We find that the local gravitational redshift effect remains negligible, provided that CMB data or some future redshift survey data are added to break the degeneracy between parameters.
\end{abstract}
\maketitle
\section{Introduction}

The cosmological principle that our universe is homogeneous and isotropic leads to the Friedmann-Lema\^{i}tre-Robertson-Walker metric
  \begin{equation}
  ds^2=-dt^2+a^2(t)\left(\frac{dr^2}{1-Kr^2}+r^2(d\theta^2+\sin^2\theta\ d\phi^2)\right), \label{metric}
  \end{equation}
  where $K$ indicates the spatial curvature, $t$ is the cosmological time, and $r$, $\theta$, $\phi$ are spherical coordinates. The scale factor $a$ represents the relative size of the universe. Without loss of generality, we normalize the scale factor today to unity.

The expansion of the universe leads to a cosmological redshift $z=1/a-1$. For an object at cosmological distance, its redshift contains two additional small components - the redshift due to the peculiar motion of the object and the redshift caused by the difference between the remote and local gravitational potentials. When averaged over many objects in a narrow redshift bin, peculiar motion and remote gravitational potential only lead to unbiased errors, while the local gravitational potential leads to a biased component $z_g$ that is typically of order a few~$\times 10^{-5}$~\cite{Local}. Such a tiny redshift correction at the first glance should be negligible in cosmological data analysis. On the contrary, Ref.~\cite{Local} showed that ignoring $z_g$ in a Type Ia supernovae (SNe) data analysis leads to percent-level biases on cosmological parameters. For modern precision cosmology, percent-level biases either in the dark energy density parameter $\Omega_\Lambda$ or in the dark energy equation of state $w$ are noticeable systematics. Ref.~\cite{Local_H} revisited this problem and found that the percent-level biases are mostly along the degeneracy direction of parameters. Showing that the biases drop significantly to typically below $0.1\sigma$ level when CMB data are added to break the parameter degeneracies, the author concluded that a typical local gravitational redshift is negligible for the data analysis of {\it currently available} SNe. This conclusion was further supported by Ref.~\cite{calcino2017need}, where an updated SNe catalog~\cite{JLA} was used.

This manuscript is a continuation of Ref.~\cite{Local_H} to study the impact of local gravitational redshift on {\it future} SNe projects. We forecast a  SNe survey based on the proposal of Wide-Field InfraRed Survey Telescope (WFIRST) project~\cite{WFIRST, WFIRST19}. To break the strong degeneracy between parameters, we use either mock data of future redshift surveys or a reduced covariance matrix from the currently available CMB data from Planck satellite~\cite{aghanim2018planck,wang2007observational,wang2013distance,zhai2018robust}. For future redshift surveys we consider a mock spectroscopic redshift survey that is similar to the Euclid project~\cite{Euclid}. For comparison we also simulate a mock photometric redshift survey in accordance to the large synoptic survey telescope (LSST) project~\cite{LSST}. We dub these mock data sets WFIRST-like, Euclid-like and LSST-like, respectively, to distinguish our work from official forecasts by collaborations of these projects. Our forecast work should be understood as qualitative estimations for the proposed major configurations of these projects. We do not attempt to explore the exact details of these projects or the most exhaustive utilization of statistical information beyond two-point statistics. 

The standard parameters used in this manuscript are the baryon density parameter $\Omega_b$, the total matter density parameter $\Omega_m$, the amplitude $A_s$ and the spectral index $n_s$ of the primordial power spectrum of curvature fluctuations, and the Hubble constant $H_0$. Unless otherwise stated, the spatial curvature parameter $\Omega_k\equiv -\frac{K}{H_0^2}$ is treated as a free parameter and the dark energy equation of state is parameterized as $w=w_0+w_a(1-a)$~\cite{chevallier2001accelerating, linder2003exploring}. 

\section{Mock data and Best-fit Finder}

We follow Ref.~\cite{OmkForecast} to simulate the WFIRST-like, Euclid-like and LSST-like mock data for a fiducial cosmology: $H_0=67.32\,\mathrm{km\,s^{-1}Mpc^{-1}}$, $\Omega_m=0.3144$, $\Omega_b=0.0494$, $A_s = 2.10\times 10^{-9}$, $n_s = 0.966$, $\Omega_k=0$, $w_0=-1$, and $w_a=0$. For the WFIRST-like mock data, an additional $z_g$ shift is added to the redshift of each supernova sample. When fitting the cosmological parameters, we ignore the $z_g$ contribution and typically find best-fit parameters that differ from the input fiducial parameters. To estimate the significance of the impact of local gravitational redshift, we compare the biases, namely, the differences between the best-fit parameters and the fiducial parameters, with the standard deviations of parameters.

We sketch below the key structures of the best-fit finder for WFIRST-like + Euclid-like mock data. For other combinations and more detailed description of the forecast techniques, the reader is referred to Ref.~\cite{OmkForecast}.

To get the biases due to the local gravitational redshift, we use Newton-Raphson method to find the best-fit parameters that render the $\chi^2$, defined as
\begin{equation}
  \chi^2=\sum_{\text{SN samples}}\frac{\left(\mu_{\rm th}-\mu_{\rm obs}\right)^2}{\sigma_{\rm \mu}^2}+ \sum_{\text{redshift bins}}\left(P_{\rm th}-P_{\rm obs}\right)^{T}\mathrm{Cov}^{-1}\left(P_{\rm th}-P_{\rm obs}\right),\label{Eq:Chi_square}
\end{equation}
reaches its minimum. The two sums on the right-hand side stand for $\chi^2$ contribution from the WFIRST-like SNe mock data and that from the Euclid-like redshift survey mock data, respectively. In the WFIRST-like $\chi^2$ term, the theoretical distance modulus $\mu_{\rm th}$ is compared with the simulated distance modulus $\mu_{\rm obs}$. The uncertainty $\sigma_\mu$ includes an intrinsic dispersion of the absolute magnitude of Type Ia supernova and contributions from gravitational lensing and the peculiar motion of supernova. We considered 17 uniform redshift bins from $z=0$ to $z=1.7$. The numbers of supernovae in each redshift bin are listed in Table~\ref{Table:SN}. We randomly generate the mock data, the redshift of each supernova, in each redshift bin. In the best-fit finder, we analytically marginalize over the nuisance calibration parameter, that is, SN magnitude at a standard distance $h^{-1}\mathrm{Mpc}$.

\begin{table}
  \caption{Number of Supernovae in Each Redshift Bin for WFIRST-like Survey\label{Table:SN}}
  \begin{center}
    \begin{tabular}{cccccccccccccccccc}
    \hline
    \hline
    $z_{\rm min}$ & 0 & 0.1 & 0.2 &0.3 &0.4 &0.5 &0.6 &0.7 &0.8 &0.9 &1.0 &1.1 &1.2 &1.3 &1.4 &1.5 &1.6 \\
    $z_{\rm max}$ & 0.1 & 0.2 &0.3 &0.4 &0.5 &0.6 &0.7 &0.8 &0.9 &1.0 &1.1 &1.2 &1.3 &1.4 &1.5 &1.6 &1.7 \\
    \hline
    Number &500 &69 &208 &402 &223 &327 &136 &136 &136 &136 &136 &136 &136 &136 &136 &136 &136 \\
    \hline
    \end{tabular}
  \end{center}
  \end{table}
    
In the Euclid-like $\chi^2$ term, $P$ is the galaxy power spectrum with redshift-space distortion~\cite{kaiser1987n}, modeled as
\begin{equation}
  P(k,\mu;z)=(b+f\cos^2\theta )^2 P_m(k) \exp \left\{ -k^2\left[R_{\parallel}^2\cos^2\theta + R_{\perp}^2\sin^2 \theta \right] \right\}+\frac{1}{\epsilon \bar{n}_{\rm obs}},\label{eq:P}
\end{equation}
where $\theta$ is the angle between the wave vector $\bm{k}$ and the line-of-sight direction. The observed galaxy density $\bar{n}_{\rm obs}$ multiplied by the fraction parameter $\epsilon$ gives the number density of galaxies with measured spectroscopic redshift. The  matter power spectrum $P_m(k)$ can be computed from cosmology with a standard Boltzmann solver~\cite{CAMB}. The linear galaxy bias $b$ is parameterized as  $b(z, k) = (b_0+b_1 z)^\beta e^{-\alpha k^2}$, where $b_0, b_1, \alpha,\beta$ are nuisance parameters. To generate Euclid-like mock data we use a fiducial bias $b = \sqrt{1+z}$, which corresponds to $b_0=b_1=1$, $\alpha=0$ and  $\beta=1/2$. The smearing lengths in line-of-sight and perpendicular directions, $R_{\parallel}$ and $R_{\perp}$, describe the nonlinear smearing of galaxy power spectrum. See Ref.~\cite{OmkForecast} for more detailed description of these quantities, as well as how the covariance matrix $\mathrm{Cov}$ are computed.

The mock Euclid-like survey covers eight uniform redshift bins from $z=0.5$ to $z=2.1$, with sky coverage 15,000 square degrees. In each redshift bin, we use conservative cutoffs of wavenumbers and include non-Gaussian corrections in the covariance matrix $\mathrm{Cov}$~\cite{OmkForecast, carron2015information}. More detailed specifications are given in Table~\ref{Table:Euclid}.
\begin{table}
  \caption{Redshift bins, number density of observed galaxies and wavenumber range for the Euclid-like mock data\label{Table:Euclid}}
  \begin{center}
    \begin{tabular}{cccc}
    \hline
    \hline
     redshift range &$\bar{n}[10^{-3}h^3\rm{Mpc^{-3}}]$ & $k_{\min}[h/\rm{Mpc}]$ &  $k_{\max}[h/\rm{Mpc}]$ \\ 
    \hline
    0.5-0.7 & $3.56$ & 0.0061 &0.09 \\
    0.7-0.9 & $2.82$ & 0.0054 &0.11 \\
    0.9-1.1 & $1.81$ & 0.0051 &0.12 \\
    1.1-1.3 & $1.44$ & 0.0048 &0.14 \\
    1.3-1.5 & $0.99$ & 0.0047 &0.16 \\
    1.5-1.7 & $0.55$ & 0.0046 &0.18 \\
    1.7-1.9 & $0.29$ & 0.0045 &0.20 \\
    1.9-2.1 & $0.15$ & 0.0045 &0.22 \\
    \hline
    \end{tabular}
  \end{center}
  \end{table}

Finally, to test the robustness of the Newton-Raphson method, we use Monte Carlo Markov Chains (MCMC) as an alternative best-fit finder algorithm. We find that the results from the two best-fit finders (Newton-Raphson and MCMC) are consistent up to two significant digits.

\section{Results}

  \begin{table}
 \caption{Biases of parameters for WFIRST-like + Euclid-like mock data are much smaller than their $1\sigma$ errors.\label{Table:bias} }
  \begin{center}
\begin{tabular}{cccc} 
  \hline
  \hline
  Parameter & $1\sigma$ error & bias ($z_g=4\times10^{-5}$)  & bias ($z_g=-4\times10^{-5}$) \\
  \hline
  $w_0$ & $2.3\times10^{-2}$&$-7.2\times10^{-3}$ & $7.3\times10^{-3}$ \\
  $w_a$ &$1.1\times10^{-1}$& $1.8\times10^{-2}$ & $-1.8\times10^{-2}$\\
  $\Omega_m$ &$1.5\times10^{-3}$ & $-2.5\times10^{-4}$ & $2.5\times10^{-4}$\\
  $\Omega_k$ & $8.5\times10^{-3}$& $-9.2\times10^{-4}$ & $9.3\times10^{-4}$ \\
  \hline
\end{tabular}
\end{center}
\end{table}

  The $1\sigma$ uncertainties (for $z_g=0$ case) and biases for typical values $z_g=\pm 4\times 10^{-5}$ are listed in Table~\ref{Table:bias}. The biases due to the local gravitational redshift are much smaller than the parameter uncertainties. The most significant case is the bias of $w_0$, which is about $0.3\sigma$.

\begin{figure*}
  \includegraphics[width=0.48\linewidth]{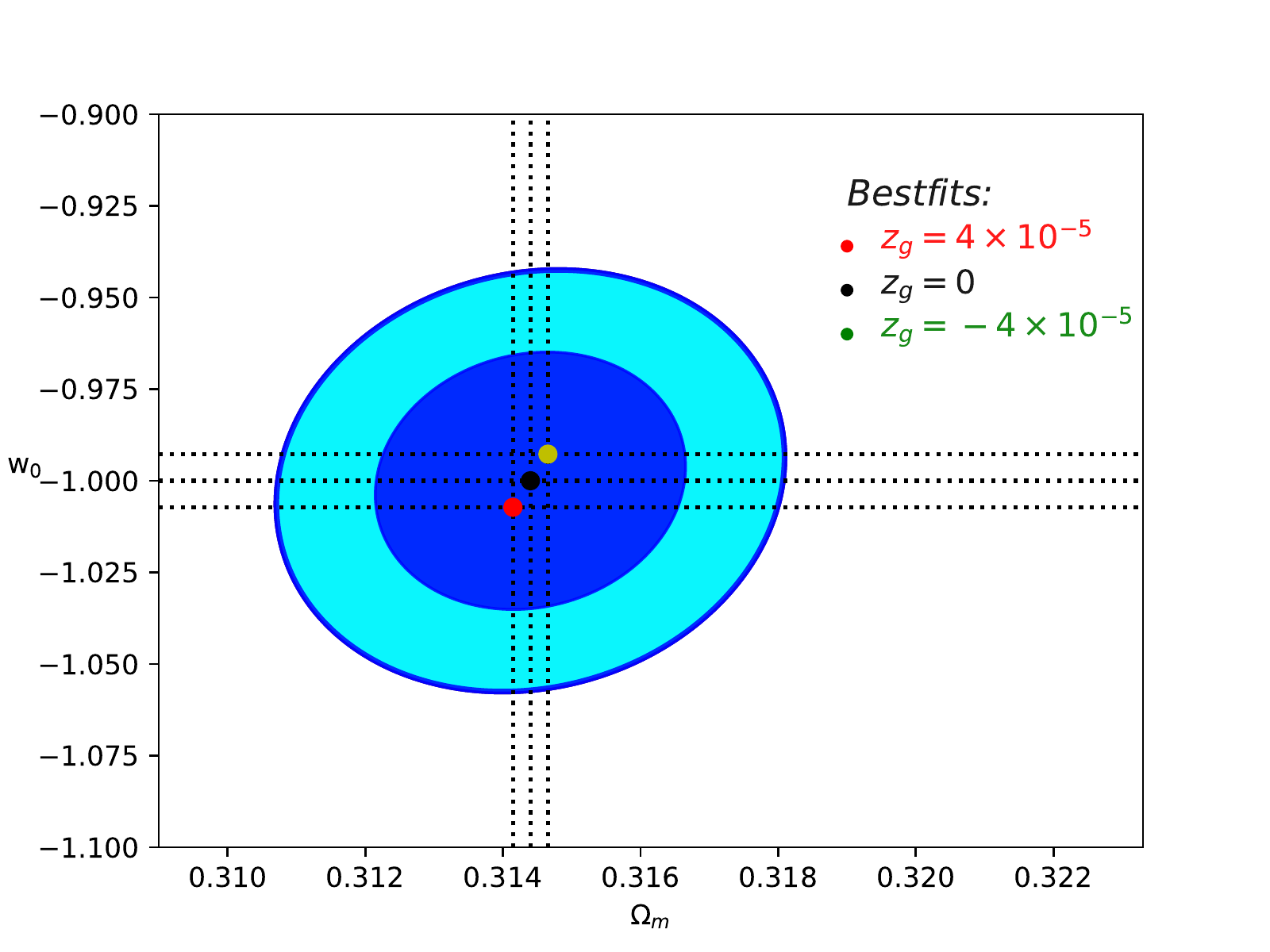}
  \includegraphics[width=0.48\linewidth]{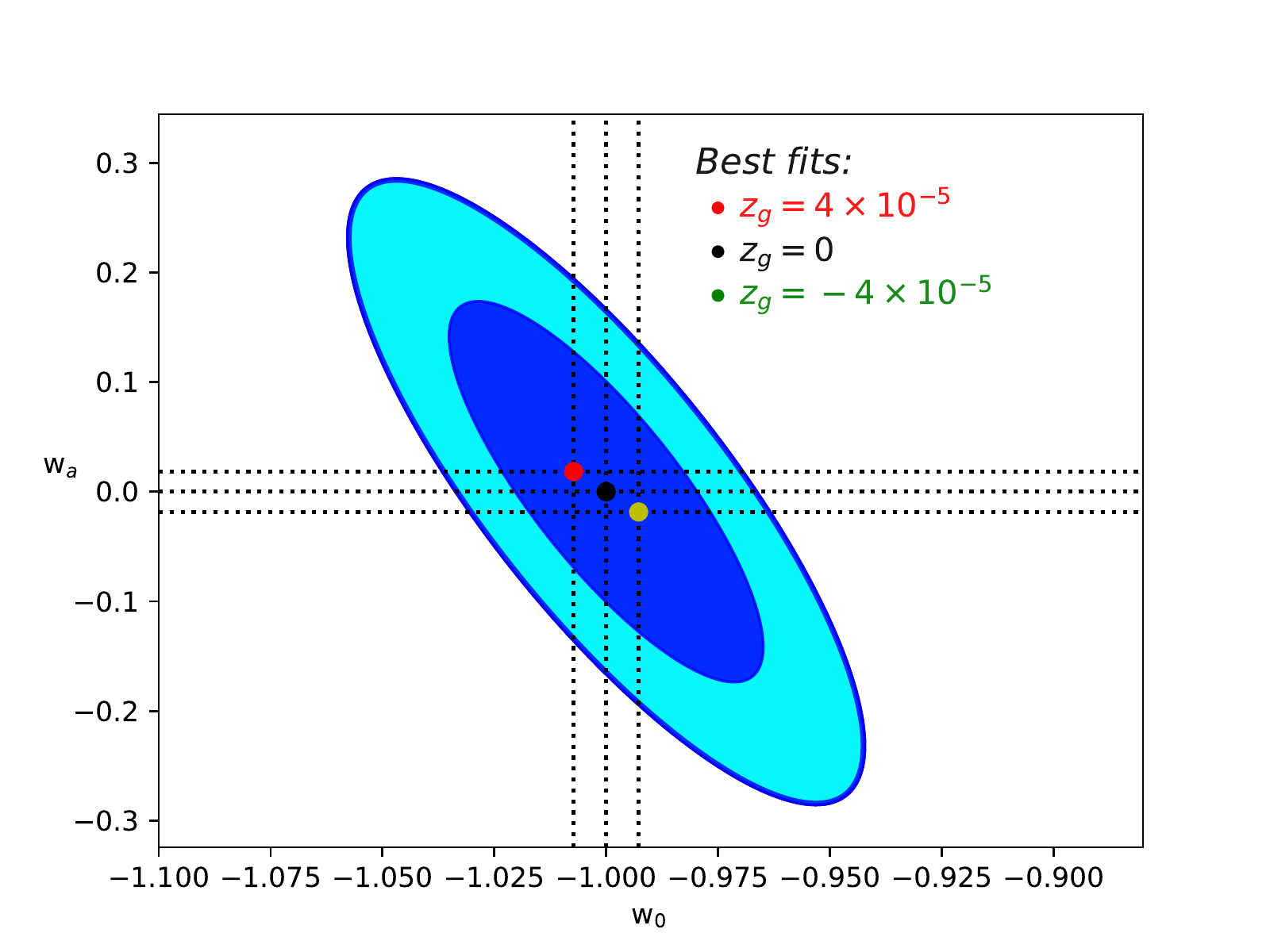}
  \caption{The best-fit parameters ($\Omega_m$ and $w_0$ in the left panel; $w_a$ and $w_0$ in the right panel) for mock data simulated with $z_g=\pm 4\times 10^{-5}$ and $z_g=0$, respectively. In all cases Euclid-like data are combined with WFIRST-like mock data to break the parameter degeneracies. The contours are 68.3\% confidence level and 95.4\% confidence level constraints for the $z_g=0$ case. \label{fig:biases}}
\end{figure*}

\begin{figure*}
  \includegraphics[width=0.48\linewidth]{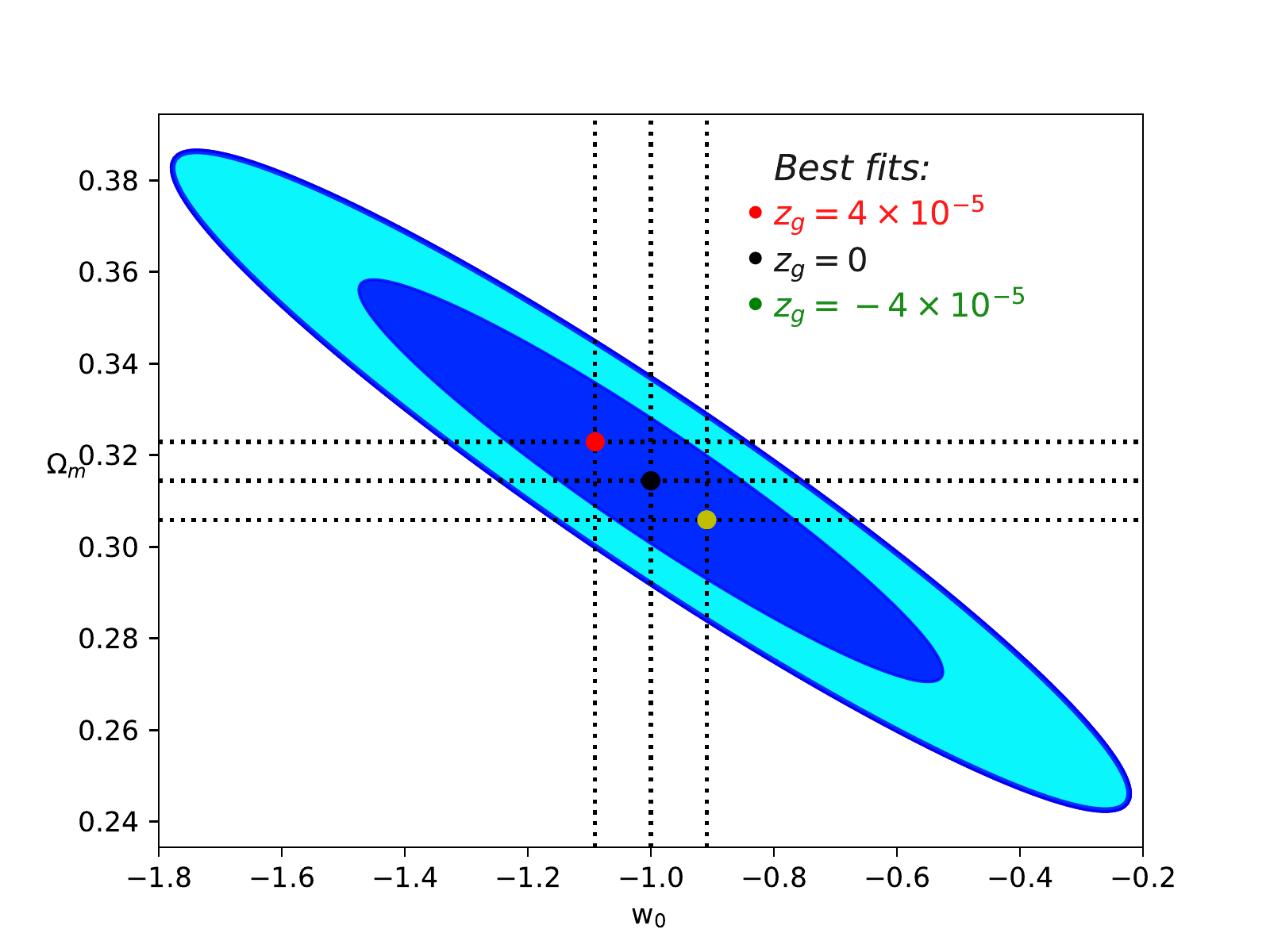}
  \caption{The best-fit parameters ($\Omega_m$ and $w_0$) for mock data simulated with $z_g=\pm 4\times 10^{-5}$ and $z_g=0$, respectively. It is obtained solely from WFIRST-like supernovae mock data. The contours are 68.3\% confidence level and 95.4\% confidence level constraints for the $z_g=0$ case. \label{fig:SN_only}}
\end{figure*}

To understand better the $0.3\sigma$ bias in $w_0$, we plot in Fig.~\ref{fig:biases} the biases against the marginalized uncertainties in projected $w_0$-$\Omega_m$ and $w_0$-$w_a$ spaces. A noticeable degeneracy between $w_0$ and $w_a$, which we interpret as the main source of the relatively significant $w_0$ bias, can be seen from the right panel.

For comparison and to show the importance of combining different cosmological data sets, we compute the biases solely from supernovae data. In this case we allow $\Omega_m$, $w_0$ and $\Omega_k$ to vary and fix $w_a=0$. (Allowing $w_a$ to vary will lead to even more significant biases.) In agreement with Refs.~\cite{Local, Local_H}, we find relatively significant biases along the degeneracy directions. One example is shown in Fig.~\ref{fig:SN_only}.

Nevertheless, when degeneracies between cosmological parameters are broken by combined data sets, the biases of parameters are typically $\sim 0.1\sigma$, which is not fatal and can be considered a secondary source of systematics in the data analysis.

Finally, we find other combinations such as WFIRST-like + LSST-like and WFIRST-like + CMB all give similar results. For the CMB data we neglected the subtle effect that the local gravitational potential also leads to an additional CMB temperature monopole. Ref.~\cite{Yoo2019} studied this effect and find that the corresponding biases of cosmological parameters are much smaller than $0.1\sigma$.

\section{Conclusions}

We studied the impact of local gravitational redshift on the future SNe observations. Even for the very general model with free spatial curvature and dynamic dark energy equation of state, as long as CMB data or some future redshift surveys are combined to break the strong degeneracy between the parameters, the biases of parameters are not very significant (typically $\sim 0.1\sigma$), at least negligible for a crude estimation.

The albeit small but still noticeable biases can be again explained by the major point of Ref.~\cite{Local_H}, that noticeable biases can only be found when there is a strong degeneracy between parameters. For the non-flat dynamic dark energy model with free $\Omega_k$, $w_0$, and $w_a$, despite the addition of redshift survey or CMB data, the degeneracy between $w_0$ and $w_a$ may still be significant, leading to an interpretation of the slightly larger bias ($\approx 0.3\sigma$ on $w_0$) we found in this work.


\end{document}